\documentclass[aps,pra,twocolumn,showpacs,superscriptaddress,floatfix]{revtex4-1}

\usepackage{amsmath}
\usepackage{amssymb}
\usepackage{bm}
\usepackage{graphicx}
\usepackage{float}
\usepackage{appendix}
\usepackage{longtable}
\usepackage{color}
\begin{document}
\title[]{Plasmon-assisted two-photon absorption\\ in a semiconductor quantum dot -- metallic nanoshell composite}
\author{Bintoro S. Nugroho}
\affiliation{Jurusan Fisika, Fakultas Matematika dan Ilmu Pengetahuan Alam, Universitas Tanjungpura, Jl. Prof. Dr. Hadari Nawawi, 78115 Pontianak, Indonesia}
\author{Alexander A. Iskandar}
\affiliation{Physics of Magnetism and Photonics Research Group, Institut Teknologi Bandung, Jl. Ganesa 10, 40132 Bandung, Indonesia}
\author{Victor A. Malyshev}
\affiliation{University of Groningen, Zernike Institute for Advanced Materials, Nijenborgh 4, 9747 AG Groningen, The Netherlands}
\affiliation{Herzen State Pedagogical University, 191186 St. Petersburg, Russia}
\author{Jasper Knoester}
\affiliation{University of Groningen, Zernike Institute for Advanced Materials, Nijenborgh 4, 9747 AG Groningen, The Netherlands}


\date{\today}

\begin{abstract}

Tho-photon absorption holds potential for many practical applications. We theoretically investigate the onset of this phenomenon in a semiconductor quantum dot -- metallic nanoshell composite subjected to a resonant CW excitation. Two-photon absorption in this system may occur in two ways: incoherent -- due to a consecutive ground-to-one-exciton-to-biexciton transition and coherent -- due to a coherent two-photon process, involving the direct ground-to-biexciton transition in the quantum dot. The presence of the nanoshell nearby the quantum dot gives rise to two principal effects:  (i) -- renormalization of the applied field amplitude and (ii) -- renormalization of the resonance frequencies and radiation relaxation rates of the quantum dot, both depending on the the quantum dot level populations. We show that in the perturbation regime, when the excitonic levels are only slightly populated, each of these factors may give rise to either suppression or enhancement of the two-photon absorption. The complicated interplay of the two determines the final effect.
Beyond the perturbation regime, it is found that the two-photon absorption experiences a drastic enhancement, which occurs independently of the type of excitation, either into the one-exciton resonance or into the two-photon resonance. Other characteristic features of the two-photon absorption of the composite, emerging from the coupling between both nanoparticles, are bistability and self-oscillations.

\end{abstract}

\pacs{
    78.67.-n,  
    73.20.Mf,  
    85.35.-p   
}

\maketitle

\section{Introduction}
\label{Introduction}
Two-photon absorption (TPA), although generally a weak effect compared to the one-photon absorption, has various practical applications, which makes it a very interesting phenomenon to study and control. The principle of using TPA processes is based on the fact that many materials, while not being transparent for radiation in the visible, are transparent in the infrared. This allows one to penetrate into the bulk with infrared light, where subsequently, through the TPA process the energy of two infrared photons may be used to trigger processes that require optical energies. 
Well-known examples of applications of this principle are microfabrication via 3D photopolymerization~\cite{MaruoOL1997,Baldacchini2016}, bioimaging~\cite{SvobodaNeuron2006}, and optical data storage~\cite{StricklerOptLett1991,CorredorAdvMat2006,MakarovJOSAB2007}. Furthermore,TPA is widely used for internal modification of bulk media (see Ref.~\cite{VerburgOE2014} and references therein) as well as for probing electronic states which are dipole forbidden due to parity.~\cite{KalitaPRA2018}. Plasmon-assisted TPA is used to improve efficiency of silicon photodetectors for optical correlators in the near-infrared~\cite{SmolyaninovOL2016} as well as to enhance the TPA in photoluminescent semiconductor nanocrystals~\cite{MarinACSPhoton2016} and fluorophores~\cite{RaborAnalyst2019}.

In this paper, we study TPA in composites that consist of a semiconductor quantum dot (SQD) and a closely spaced metal nanoparticle (MNP). It is well established that the presence of a MNP nearby a SQD strongly affects the optical response of the SQD as a consequence of the polarizability of the MNP. Notable phenomena that have been studied in this context are: bistable optical response~\cite{ArtusoNL2008,ArtusoPRB2010,MalyshevPRB2011,LiOE2012,NugrohoJCP2013}, linear and nonlinear Fano resonances~\cite{ZhangPRL2006,KosionisJPCC2012,NugrohoPRB2015}, gain without inversion~\cite{SadeghiNanotechnology2010}, and several other effects~\cite{SadeghiPRB2009,AntonPRB2012,NugrohoJOpt2017}.
In a recent paper~\cite{NugrohoPRB2019}, we have studied theoretically two-photon Rabi oscillations (TPRO) in a SQD-MNP composite and found a significant influence of the SQD-MNP coupling on the TPRO. Here, we show that also the TPA of a SQD may be influenced strongly by the presence of a nearby MNP. As in~\cite{NugrohoPRB2019}, we adopt for the SQD a ladder-like three-level model which includes ground, one-exciton, and biexciton states. For the MNP, we consider a metallic nanoshell (MNS), a spherical nanoparticle consisting of a dielectric core covered by a thin metallic layer (usually gold). MNSs are best-known in relation to their usage in cancer therapy~\cite{LooTechnolCancerResTreat2004} and bioimaging~\cite{LooOptLett2005}. From the viewpoint of optical applications, MNSs are of great interest due to their high spectral tunability originating from plasmon hybridization of the inner and outer surface of the metallic shell~\cite{ProdanNL2003,HarrisBG2008}. The hybridization gives rise to two plasmon resonances. The lower-energy one couples strongly to incident light, whereas the higher-energy one is anti-bonding and therefore weakly interacts with light. Thus, MNSs are ideal partners for combination with quantum emitter to resonantly enhance the optical response of the latter.

The present study is focused on exploring the plasmonic effect on the TPA of a SQD-MNS composite. As an example, we choose an InGaAs/GaAs SQD, absorbing in the infrared, in close proximity to an Au-silica MNS tuned in resonance with the SQD excitonic transitions. We show that the SQD-MNS coupling strongly affects the TPA of the composite as compared to an isolated SQD, resulting in bistability, self-oscillations, and a drastic enhancement of the TPA within a certain range of the external field magnitude. 

This paper is organized as follows. In the next section, we present the model system and the mathematical formalism for its description. In Sec.~\ref{Pertubative treatment}, the perturbation theory is used to study the TPA and the effects of the presence of a MNS nearby the SQD on the TPA (renormalization of the external field magnitude and exciton energies and relaxation rates) are explored.
In Sec.~\ref{Numerical results}, we report the results of numerical calculations of the TPA, extending also beyond the perturbation regime, for a set of parameters characteristic for an InGaAs/GaAs SQD -- Au-silica MNS composite and discuss these. Section~\ref{Summary} summarizes the paper. In the Appendix, an exact parametric method of solving the nonlinear steady-state problem is described.

\section{Modeling the SQD-MNS composite}
\label{Theoretical Model}
\begin{figure*}[ht]
\includegraphics[width=0.8\linewidth]{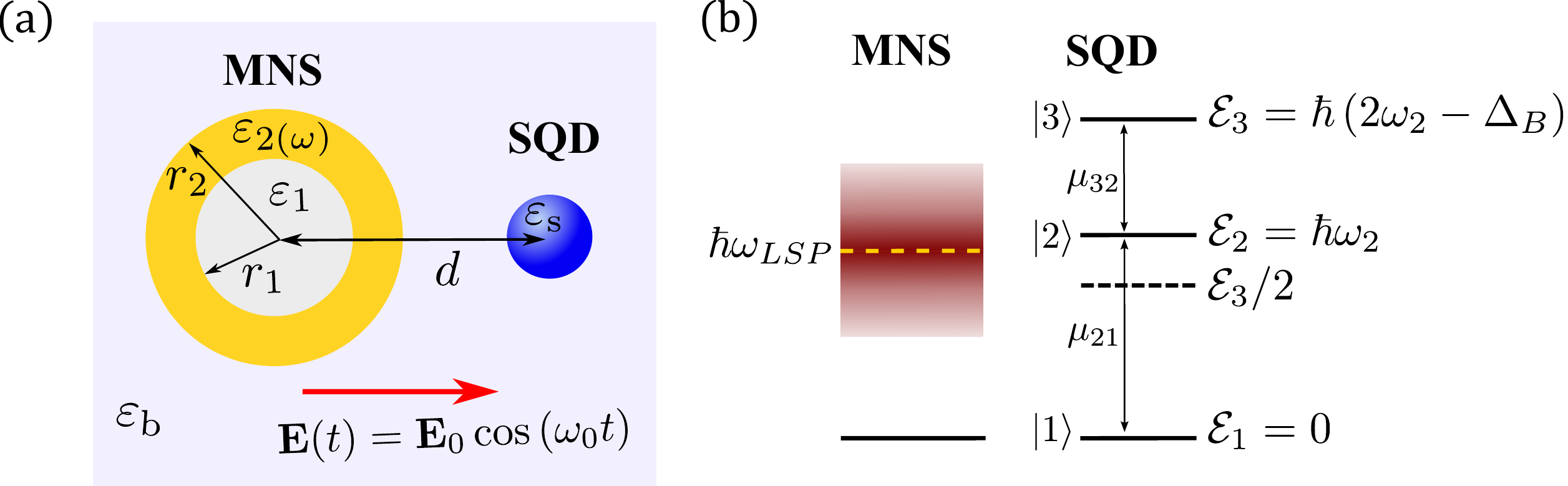}
\caption{\label{Schematics}(a)~Schematics of a SQD-MNS composite subjected to a monochromatic field $\mathbf{E}(t) = \mathbf{E}_0 \cos (\omega_0 t)$ linearly polarized along the system's axis (indicated by the red arrow). The MNS core of radius $r_1$ represents a dispersionless dielectric with the dielectric constant $\varepsilon_1$. The shell's metallic layer of  thickness $r_2 - r_1$ is characterized by the dielectric function $\varepsilon_2(\omega)$. $\varepsilon_s$ denotes the SQD dielectric constant. The two nanoparticles are separated by their center-to-center distance $d$ and embedded in a dispersionless isotropic medium with permittivity $\varepsilon_b$. (b)~Energy diagrams of the MNS (left) and the ladder-type three-level SQD (right). The excited state of the MNS represents a broad continuum centered at the frequency of the LSP's resonance, $\omega_\mathrm{LSP}$ (shown by the dashed yellow line). For the SQD, $|1 \rangle$, $|2 \rangle$, and $|3 \rangle$ are the ground, one-exciton, and biexciton  states, respectively. The energies of these states are $\mathcal{E}_1 =0$, $\mathcal{E}_2 =\hbar \omega_2$ and $\mathcal{E}_3 = 2\hbar(\omega_2 -\Delta_B/2)$, where $\hbar\Delta_B$ is the biexciton  binding energy. Allowed transitions with the corresponding transition dipole moments $\boldsymbol{\mu}_{21}$ and $\boldsymbol{\mu}_{32}$ are indicated by the solid double-headed arrows. The dashed black line shows the location of the coherent two-photon resonance $\omega_3/2 = \omega_2 - \Delta_B/2$ (with simultaneous absorption of two photons).
}
\centering
\end{figure*}

The geometry of our system is shown in Fig.~\ref{Schematics}(a). We consider a heterodimer comprising a SQD and a closely spaced MNS subjected to a monochromatic field $\mathbf{E}(t) = \mathbf{E}_0 \cos (\omega_0 t)$ of amplitude $\mathbf{E}_0$ and frequency $\omega_0$, polarized along the system's axis. The MNS consists of a core of radius $r_1$, representing a dispersionless dielectric with the dielectric constant $\varepsilon_1$, and a metallic layer (covering the core) of thickness $r_2 - r_1$ and with dielectric function $\varepsilon_2(\omega)$. The dielectric properties of the SQD are characterized by the dielectric constant $\varepsilon_s$. The SQD and MNS are separated by a center-to-center distance $d$ and embedded in a dispersionless isotropic medium with permittivity $\varepsilon_b$. We assume the system's size small compared to the optical wavelength, a condition that holds for the parameters used in our study. This allows one to apply the quasistatic approximation and neglect retardation effects.

\subsection{MNS}
\label{MNS}
Figure~\ref{Schematics}(b) (left) shows the level diagram of the MNS. The resonant incident field excites localized surface plasmons (LSPs) in the metal. In the case of a MNS, the metallic layer, covering the dielectric core, supports two plasmon resonances corresponding to the inner and the outer surface of the layer. If the layer is thin enough, the resonances strongly interact with each other, giving rise to two new modes, a bright and a dark one. The frequency of the former (latter) is shifted down (up) with respect to the bare position~\cite{ProdanNL2003}. The shift is highly sensitive to the layer thickness which results in a broad-band tunability of the MNS's bright plasmon resonance across the visible and the near infrared~\cite{HarrisBG2008}. Within the classical approach, the MNS optical response is well described by the MNS's frequency dependent polarizability $\alpha_1(\omega)$. In the quasistatic limit,  $\alpha_1(\omega)$ is given by~\cite{BohrenBook1983}
\begin{widetext}
\begin{equation}
  \alpha_1(\omega) = 4\pi r_2^{3}
                 \frac{\left[\varepsilon_1 + 2 \varepsilon_2(\omega) \right][\varepsilon_2(\omega)
                 -\varepsilon_b]+ \left(r_1/r_2 \right) ^{3}\left[\varepsilon_1-
                 \varepsilon_2(\omega) \right]
                 \left[\varepsilon_b +2\varepsilon_2(\omega) \right]}
                 {\left[\varepsilon_2(\omega) + 2\varepsilon_b\right) \left[ \varepsilon_1
                 + 2\varepsilon_2(\omega) \right]+2\left( r_1/r_2 \right)^{3}
                 \left[\varepsilon_2(\omega)-\varepsilon_b \right]
                 \left[\varepsilon_1-\varepsilon_2(\omega)\right]}~.
\label{alpha_1}
\end{equation}
\end{widetext}
Equation~(\ref{alpha_1}) is valid for MNS sizes $2r_2$ small compared to the wavelength of the incident field, $2\pi c/\omega_0$, $c$ being speed of light. For the infrared-to-visible range of wavelengths, this limits the MNS size to $\lesssim 100$ nm~\cite{Maier2007}. The lower bound for $2r_2$ is dictated by quantum size effects, coming into play for $r_2 \lesssim 5$ nm~\cite{SchoNature2012}. In our study, we consider MNS sizes for which Eq.~(\ref{alpha_1}) may safely be applied.

It is apparent from equation (\ref{alpha_1}) that $\alpha_1(\omega)$ experiences resonant enhancement when the absolute value of the denominator in Eq.~(\ref{alpha_1}) reaches its minimum (Fr\"{o}hlich resonance condition). The latter determines the frequency of the LSP resonance, $\omega_{LSP}$. Accordingly, the plasmonic states of the MNS constitute a ground state and a broad continuum of excited states, as shown in Fig.~\ref{Schematics}(b) (left). The LSP resonance is shown by the dashed yellow line.

\begin{figure}
\includegraphics{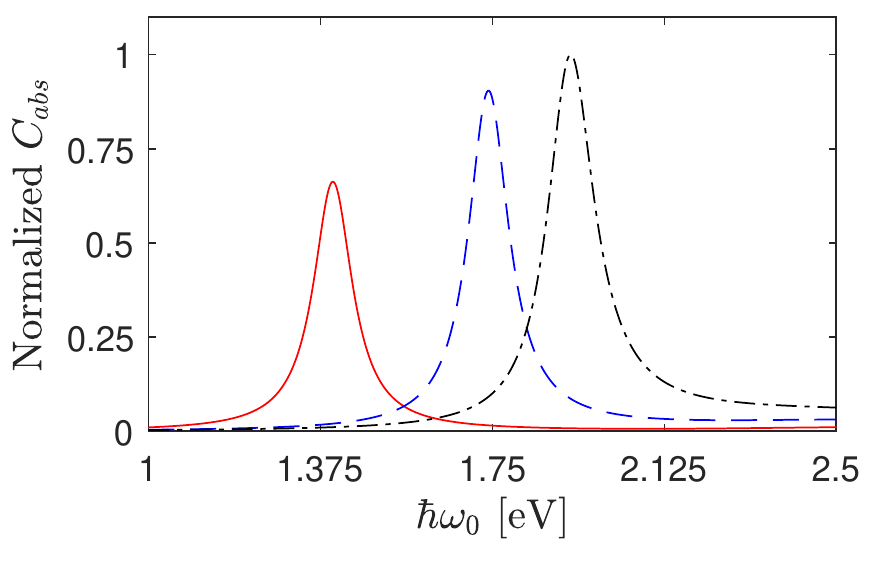}
\caption{\label{SpectMNP} Absorption cross-section $C_{abs}(\omega)$  of the silica--Au MNS with a silica core of $r_1 = 9$~nm, calculated from Eq.~(\ref{alpha_1}) for three different shell thickness: $r_2 - r_1 = 1~\mathrm{nm}$ (solid curve), $r_2 - r_1 = 2~\mathrm{nm}$ (dashed curve), and $r_2 - r_1 = 3~\mathrm{nm}$ (dash-dotted curve). 
}
\centering
\end{figure}

Throughout the paper, we will consider silica-Au core-shell nanoparticles embedded in a silica host, taking, accordingly, $\varepsilon_1 = \varepsilon_b = 2.16$, while calculating the gold dielectric function $\varepsilon_2(\omega)$ by means of the modified Drude model~\cite{DerkachovaPlasmon2016}. To illustrate the strong sensitivity of the MNS plasmon resonance $\omega_{LSP}$ to the geometrical parameters of the MNS, we present in Fig.~\ref{SpectMNP} the results of calculations of the MNS absorption cross-section $C_{abs}(\omega) \propto \mathrm{Im}[\alpha(\omega)]$, keeping the core radius $r_1 = 9~\mathrm{nm}$ fixed and varying the outer shell radius $r_2$. As is seen from the figure, changing the shell thickness from 1 to 3~$\mathrm{nm}$ dramatically affects the location of the MNS plasmon resonance, moving it from the infrared to the visible upon increasing the shell thickness.

\subsection{SQD}
\label{SQD}
Figure~\ref{Schematics}(b) (right) shows the level diagram and allowed transitions of the SQD. The optical excitations in the SQD are excitons. In such a system, the degenerate one-exciton state is split into two linearly polarized one-exciton states due to the anisotropic electron-hole exchange interaction~\cite{StuflerPRB2006,JundtPRL2008,GerardotNJP2009}. In this case, the ground state is coupled to the biexciton state via the linearly polarized one-exciton state. Thus, the system effectively acquires a three-level ladder-like structure with the ground ($|1\rangle$), one-exciton ($|2\rangle$), and biexciton ($|3\rangle$) state, as shown in  Fig.~\ref{Schematics}(b) (right). The energies of these states are $\mathcal{E}_1=0$, $\mathcal{E}_2=\hbar\omega_{2}$, and $\mathcal{E}_3=\hbar\omega_3 = 2\hbar(\omega_2 - \Delta_B/2)$, respectively, where $\hbar\Delta_B$ is the biexciton binding energy. Within this scheme, the allowed transitions induced by the applied field are $|1\rangle \leftrightarrow |2\rangle$ and $|2\rangle \leftrightarrow |3\rangle$ with corresponding transition dipole moments $\boldsymbol{\mu}_{21} (= \boldsymbol{\mu}_{12})$ and $\boldsymbol{\mu}_{32} (= \boldsymbol{\mu}_{23})$, accordingly. The transition between the ground state $|1\rangle$ and biexciton state $|3\rangle$ is dipole-forbidden by parity and can only be achieved by the simultaneous absorption of two photons.

The optical dynamics of the SQD is described by means of the Lindblad quantum master equation for the density operator $\rho(t)$, which in the rotating (with frequency $\omega_0$ of the applied field) frame reads~\cite{Lindblad1976,Blum2012}
\begin{widetext}
\begin{subequations}
\label{MasterEqAndHamiltonian}
\begin{equation}
\label{DensityMasterEq}
\dot{\rho}(t) = -\frac{i}{\hbar} \left[\mathcal{H}^{\mathrm{RWA}}(t),\rho(t)\right ]
           + \mathcal{L}_{\gamma}\{\rho(t)\} + \mathcal{L}_{\Gamma}\{\rho(t)\}~,
\end{equation}
\begin{equation}
\label{HamiltonianRWA}
\mathcal{H}^\mathrm{RWA}(t) =  \hbar \left( \Delta_{21}\sigma_{22} + \Delta_{31} \sigma_{33}\right)
        -  \hbar \left[ {\Omega}_{21}(t)\sigma_{21} + \Omega_{32}(t) \sigma_{32} + \mathrm{H.c.}\right ]~,
\end{equation}
\begin{equation}
\label{Lindblad_rel}
\mathcal{L}_{\gamma}\{\rho(t)\} = \frac{\gamma_{21}}{2} \left( \left[ \sigma_{12} \rho(t),
           \sigma_{21} \right]  + \left[ \sigma_{12},\rho(t)\, \sigma_{21} \right]\right)
           + \frac{\gamma_{32}}{2} \left( \left[ \sigma_{23} {\rho}(t),\sigma_{32}\right]
           + \left[ \sigma_{23},\rho(t)\, \sigma_{32} \right]\right)~,
\end{equation}
\begin{equation}
\label{Lindblad_dep}
\mathcal{L}_{\Gamma}\{\rho(t)\} = \Gamma_{2} \left( \left[ \sigma_{22} \rho(t),
           \sigma_{22} \right]  + \left[ \sigma_{22},\rho(t)\, \sigma_{22} \right]\right)
           + \Gamma_{3} \left( \left[ \sigma_{33} {\rho}(t),\sigma_{33}\right]
           + \left[ \sigma_{33},\rho(t)\, \sigma_{33} \right]\right)~.
\end{equation}
\end{subequations}
\end{widetext}
Here, $\mathcal{H}^{\mathrm{RWA}}(t)$ is the SQD Hamiltonian in the rotating frame, $[\mathrm{A,B}]$ denotes the commutator, $\mathcal{L}_{\gamma}\{\rho(t)\}$ is the Lindblad operator describing the radiation relaxation of the SQD states $|2\rangle$ and $|3\rangle$ with  constants $\gamma_{21}$ and $\gamma_{32}$, respectively, while $\mathcal{L}_{\Gamma}\{\rho(t)\}$ accounts for dephasing of the states $|2\rangle$ and $|3\rangle$ with rates $\Gamma_2$ and $\Gamma_3$, respectively, and $\sigma_{ij}=|i\rangle \langle j|$ ($i,j = 1,~2,~3$).
In Eq.~(\ref{HamiltonianRWA}), $\hbar\Delta_{21} = \hbar(\omega_2 - \omega_0)$ and $\hbar(\Delta_{31} = \hbar(\omega_3 - 2\omega_0)$ are the energies of states $|2 \rangle$ and $|3 \rangle$ in the rotating frame, respectively. $\Omega_{21}(t) = \boldsymbol{\mu}_{21}\cdot\boldsymbol{E}_{SQD}(t)/(2\hbar)$ and $\Omega_{32}(t) = \boldsymbol{\mu}_{32}\cdot \boldsymbol{E}_{SQD}(t)/(2\hbar)$ are the slowly varying Rabi amplitudes of $\boldsymbol{E}_{SQD}(t)$ for the corresponding transitions, where $\boldsymbol{E}_{SQD}(t)$ is the amplitude of the field acting on the SQD.

For the sake of simplicity, we assume that the transition dipoles are parallel to each other ($\boldsymbol{\mu}_{32}=\mu \boldsymbol{\mu}_{21}$) and to the acting field as well. Then $\gamma_{32}=\mu^2 \gamma_{21} \equiv \mu^2 \gamma$, $\Omega_{32} = \mu \Omega_{21} \equiv \mu \Omega$, and  all vectorial quantities can be considered as scalars. Finally, the system of equations for the density matrix elements $\rho_{ij}(t) = \langle i|\rho(t)|j\rangle$ takes the form
\begin{widetext}
\begin{subequations}
\label{all_dR1}
\begin{equation}
\dot{\rho}_{11} = \gamma \rho_{22} + i(\Omega^* \rho_{21} - \Omega \rho^*_{21})~,
\label{rho11}
\end{equation}
%
\begin{equation}
\dot{\rho}_{22} = - \gamma \rho_{22} + \mu^2 \gamma \rho_{33}
                + i (\Omega \rho^*_{21} - \Omega^* \rho_{21}
                + \mu \Omega^* \rho_{32} - \mu \Omega \rho^*_{32})~,
\label{rho22}
\end{equation}
\begin{equation}
\dot{\rho}_{33} = -\mu^2 \gamma \rho_{33} + i \mu \left(\Omega \rho^*_{32}
                -\Omega^* \rho_{32} \right)~,
\label{rho33}
\end{equation}
\begin{equation}
\dot{\rho}_{21} = -\left( i \Delta_{21} + \frac{1}{2} \gamma + \Gamma_2 \right)
                \rho_{21} + i ( \mu \Omega^* \rho_{31} - \Omega Z_{21} )~,
\label{rho21}
\end{equation}
\begin{equation}
\dot{\rho}_{32} = -\left[ i\Delta_{32} + \frac{1}{2} \left(1 + \mu^2 \right)\gamma 
                + \Gamma_2 + \Gamma_3 \right] \rho_{32}
                - i ( \Omega^* \rho_{31} + \mu \Omega Z_{32} )~,
\label{rho32}
\end{equation}
\begin{equation}
\dot{\rho}_{31} = -\left( i\Delta_{31} + \frac{1}{2} \mu^2 \gamma
                  + \Gamma_3 \right) \rho_{31}
                  + i (\mu \Omega \rho_{21} - \Omega \rho_{32})~,
\label{rho31}
\end{equation}
\end{subequations}
\end{widetext}
where $\Delta_{32} = \omega_3 - \omega_2 - \omega_0$ is the detuning away from the $|3 \rangle \leftrightarrow |2 \rangle$ transition and $Z_{ji} = \rho_{jj}-\rho_{ii}$ stand for the population difference between the states $|j\rangle$ and  $|i\rangle$. In Eqs.~(\ref{rho33})--(\ref{rho31}), we suppressed the time dependence of all dynamic variables.

Now, we address the Rabi amplitude $\Omega$ of the field acting on the SQD. This field consists of the applied field $\boldsymbol{E}_0$ and the field produced by the MNS at the position of the SQD. Taking into account the contribution of higher multipoles, the amplitude of the total field experienced by the SQD reads as~\cite{YanPRB2008,ArtusoPRB2011,NugrohoJOpt2017}
\begin{widetext}
\begin{equation}
\label{E_SQD}
  E_{\mathrm{SQD}} = \frac{1}{\varepsilon_s'}\left[ 1
                     + \frac{\alpha_1(\omega_0)}{2 \pi d^3} \right]E_0
                     + \frac{1}
                     {16 \pi^2 \varepsilon_0 \varepsilon_b \varepsilon_s'}
                    \sum_n \frac{n(n+1)(n+1)^2}{2} \frac{\alpha_n(\omega_0)}{d^{2n+4}} P_{\mathrm{SQD}}~,
\end{equation}
\end{widetext}
where $\varepsilon_s' = (\varepsilon_s + 2 \varepsilon_b)/(3\varepsilon_b)$ is the effective dielectric constant of the SQD, $\alpha_n(\omega)$ is the MNS's multipolar polarizability of $n$th order ($n = 1,2,3,....$) given by the expression~\citep{NaeimiJOSAB2019}
\begin{widetext}
\begin{equation}
\label{alpha_n}
 \alpha_n(\omega) = 4\pi r_2^{2n+1}
                 \frac{\left[\varepsilon_1 + \frac{n+1}{n} \varepsilon_2(\omega) \right]
                 \left[\varepsilon_2(\omega) - \varepsilon_b\right]+
                 \left( \frac{r_1}{r_2} \right)^{2n+1}\left[\varepsilon_1-\varepsilon_2(\omega)\right]
                 \left[\varepsilon_b +\frac{n+1}{n}\varepsilon_2(\omega) \right]}
                 {\left[\varepsilon_2(\omega) + \frac{n+1}{n}\varepsilon_b\right] \left[ \varepsilon_1
                 + \frac{n+1}{n}\varepsilon_2(\omega)\right]+
                 \frac{n+1}{n}\left( \frac{r_1}{r_2} \right)^{2n+1}
                 \left[\varepsilon_2(\omega)-\varepsilon_b \right]
                 \left[\varepsilon_1-\varepsilon_2(\omega)\right]}~,
\end{equation}
\end{widetext}
and $P_{\mathrm{SQD}}$ is the SQD's dipole moment amplitude defined as
\begin{equation}
\label{P_SQD}
  P_{\mathrm{SQD}} = \mu_{21} \left( \rho_{21} + \mu \rho_{32} \right)~.
\end{equation}

As may be inferred from the first term in Eq.~(\ref{E_SQD}), the applied field $E_0$ experiences renormalization (enhancement or suppression, see below) due to the presence of the nearby MNS, which is described by the second term in the square brackets. This originates from the field generated by the oscillating plasmons in the MNS. Finally, the last term in Eq.~(\ref{E_SQD}) represents the electromagnetic self-action of the SQD via the MNS: the field acting on the SQD depends on its own dipole moment $P_\mathrm{SQD}$.

Based on the above, the Rabi amplitude $\Omega = \mu_{21}E_\mathrm{SQD}/(2\hbar)$ is expressed as follows:
\begin{equation}
        \Omega = \widetilde{\Omega}_0 + G \left( \rho_{21} + \mu \rho_{32} \right) \ ,
\label{Omega}
\end{equation}
where
\begin{equation}
        \widetilde{\Omega}_0 \equiv \widetilde{\Omega}_{21}^0 = \frac{1}{\varepsilon_s'}\left[ 1
                     +\frac{\alpha_1(\omega_0)}{2 \pi d^3} \right]\Omega_{0}~,
\label{Omega_tilde}
\end{equation}
with $\Omega^0 = \mu_{21} E_0/(2\hbar)$ being the Rabi amplitude of the applied field for the $|1\rangle \leftrightarrow |2\rangle$ transition and
\begin{equation}
        G = \frac{\mu_{21}^2}{16 \pi^2 \hbar \varepsilon_0
            \varepsilon_b \varepsilon_s'} \sum_n \frac{n(n+1)(n+1)^2}{2} \frac{\alpha_n(\omega_0)}
            {d^{2n+4}}~.
\label{G}
\end{equation}
The complex-valued quantity $G = G_R + iG_I$ represents the feedback parameter, describing the self-action of the SQD via the MNS.~\cite{ArtusoNL2008,ArtusoPRB2010,MalyshevPRB2011,LiOE2012,NugrohoJCP2013}. It combines all properties of the materials and the geometry of the constituents, the contribution of higher multipoles, and it drives the nonlinear SQD-MNS's response.

The essential effects of the SQD self-action  can be uncovered after substituting Eq.~(\ref{Omega}) into Eqs.~(\ref{rho21}) and (\ref{rho32}). Doing so, one obtains
\begin{widetext}
\begin{subequations}
\begin{eqnarray}
\label{rho21extended}
\dot{\rho}_{21} = &-& \left[ i \left(\Delta_{21} + G_R Z_{21} \right)
                + \frac{1}{2}\gamma + \Gamma_2 - G_I Z_{21} \right] \rho_{21}
\nonumber\\
               &+& i ( \mu \widetilde{\Omega}_0^* \rho_{31}
                - \widetilde{\Omega}_0 Z_{21} )
                + i \left[ (\mu G^* \rho^*_{21} + \mu^2 G^* \rho^*_{32}) \rho_{31}
                - \mu G \rho_{32} Z_{21} \right]~,
\end{eqnarray}
\begin{eqnarray}
\label{rho32extended}
\dot{\rho}_{32} = &-& \left[ i(\Delta_{32} + \mu^2 G_R Z_{32})
                + \frac{1}{2} (1 +\mu^2)\gamma +  \Gamma_2 + \Gamma_3 - \mu^2 G_I Z_{32} \right]
                \rho_{32}
\nonumber\\
                &-& i ( \widetilde{\Omega}_0^* \rho_{31} + \mu \widetilde{\Omega}_0 Z_{32} )
                - i \left[ (G^* \rho^*_{21} + \mu G^* \rho^*_{32}) \rho_{31} + \mu G \rho_{21} Z_{32} \right]~.
\end{eqnarray}
\end{subequations}
\end{widetext}
As compared with an isolated SQD ($G=0$), these equations contain additional nonlinear terms. Two of these should get special attention, namely
(i) - renormalization of the SQD transition frequencies, $\Delta_{21} \rightarrow \Delta_{21} + G_R Z_{21}$ and  $\Delta_{32} \rightarrow \Delta_{32} + \mu^2 G_R Z_{32}$, and (ii) - renormalization of the damping rates of the off-diagonal density matrix elements, $\gamma/2 + \Gamma_{2} \rightarrow \gamma/2 + \Gamma_{2} - G_I Z_{21}$ and
$(1 + \mu^2)\gamma + \Gamma_{2} + \Gamma_3 \rightarrow (1 + \mu^2)\gamma  + \Gamma_2 + \Gamma_3 - \mu^2 G_I Z_{32}$, both depending on the corresponding population differences. As will be shown below, these two effects are essential in the formation and understanding of the complicated optical response of the composite.

\section{Perturbation treatment}
\label{Pertubative treatment}
\begin{figure}
\includegraphics[width=0.85 \columnwidth]{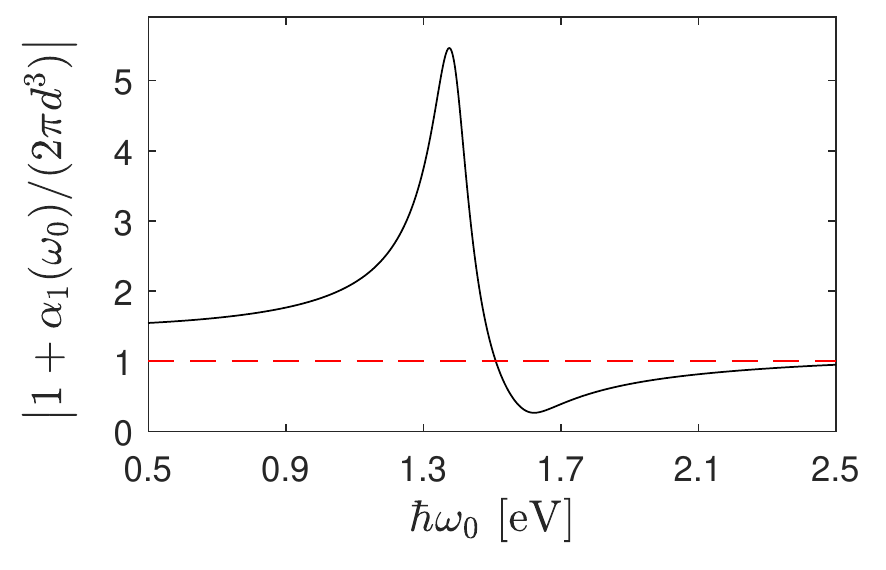}
\caption{\label{Enhancement} Frequency dependence of the modulus factor $|1 + \alpha_1(\omega_0)/(2\pi d^3)|$ for the Au-silica MNS with $r_1 = 9~\mathrm{nm}$ and $r_2 = 10~\mathrm{nm}$, as follows from Eq.~(\ref{alpha_1}).}
\centering
\end{figure}
Prior to studying the general case of arbitrary external field magnitude $|\Omega_0|$, we briefly consider the low-field limit ($|\Omega_0| \ll \Delta_B/2$) where the perturbation approach is applicable. This will help us to explicitly explore the effects of the SQD-MNS interaction on the TPA. At $|\Omega_0| \ll \Delta_B/2$, the rate $W_\mathrm{TPA}$ of the coherent TPA ($\Delta_{21} = \Delta_B/2$) is given by the second order perturbation formula
\begin{widetext}
\begin{eqnarray}
\label{W_TPA}
      W_\mathrm{TPA} = 2\pi \left| \frac{\widetilde{\Omega}_{21}^0\widetilde{\Omega}_{32}^0}{i \left(\frac{1}{2}\Delta_B - G_R \right)
                + \frac{1}{2}\gamma + \Gamma_2 + G_I} \right|^2 \frac{1}{\frac{1}{2}\mu^2\gamma + \Gamma_3} = \nonumber\\
                2\pi \frac{\mu^2}{(\varepsilon_s')^4} \left| 1 + \frac{\alpha_1(\omega_0)}{2 \pi d^3} \right|^4
                     \frac{\Omega_0^4}{ \left(\frac{1}{2}\Delta_B - G_R \right)^2
                + \left( \frac{1}{2}\gamma + \Gamma_2 + G_I \right)^2} \, \frac{1}{\frac{1}{2}\mu^2\gamma + \Gamma_3}~,
\end{eqnarray}
\end{widetext}
where $\widetilde{\Omega}_{21}^0 \equiv \widetilde{\Omega}_0$ is taken from Eq.~(\ref{Omega_tilde}) and $\widetilde{\Omega}_{32}^0 = \mu\widetilde{\Omega}_0$. Note that in our case, the intermediate state for the TPA is the one-exciton state $|2 \rangle$, which, due to the SQD-MNS interaction, is shifted in energy and broadened by the amounts $\hbar G_R$ and $\hbar G_I$, respectively (see the discussion at the end of the preceding section). This determines the denominator in Eq.~(\ref{W_TPA}). The last multiplier in Eq.~(\ref{W_TPA}) represents the density of the final states.

The modulus factor as a function of frequency, calculated by means of Eq.~(\ref{alpha_1}) for the MNS with $r_1 = 9~\mathrm{nm}$ and $r_2 = 10~\mathrm{nm}$, is depicted in Fig.~\ref{Enhancement}. As follows from the figure, depending on $\omega_0$, this factor can be both larger and smaller than unity, thus yielding, respectively, either enhancement or suppression of the TPA rate.

Also, the effect of renormalization of the energetic and relaxation characteristics of the one-exciton state on the TPA may be an enhancement or suppression of the TPA; this depends on the relationship between the constants of an isolated SQD ($\Delta_B/2, (1/2)\mu^2\gamma + \Gamma_2$) and the SQD-MNS coupling ($G_R$ and $G_I$). Enhancement occurs for $G_R \approx \Delta_B/2, G_I \ll (1/2)\mu^2\gamma + \Gamma_3$, while suppression takes place if $|G_R| \gg \Delta_B/2, G_I \gg (1/2)\mu^2\gamma + \Gamma_3$.

Summarizing, the complicated interplay of the two underlined factors determine the final effect of the MNS on the TPA of the composite (enhancement or suppression).

\section{Numerical results}
\label{Numerical results}
In what follows, we analyze the effect of the SQD-MNS coupling on the TPA of the composite beyond the perturbation regime. 
Recall that the direct ground--to--biexciton transition is dipole-forbidden. It can be achieved either via consecutive $|1\rangle \rightarrow |2\rangle \rightarrow |3\rangle$ transitions or via simultaneous absorption of two-photons of energy close to $\mathcal{E}_2 - \hbar\Delta_B/2$.

In our numerical calculations, we use parameters typical for an isolated InGaAs/GaAs quantum dot~\cite{StuflerPRB2006,GerardotNJP2009}, which absorbs light in the infrared. More specifically, the energies of the one-exciton and biexciton transitions are, respectively, $\mathcal{E}_2 = 1.34~\mathrm{eV}$ and $\mathcal{E}_3 = 2{\mathcal{E}}_2-\hbar\Delta_B$ with $\hbar\Delta_B = 2.75~\mathrm{meV}$, and the radiation decay constants of the corresponding transitions are $\hbar\gamma_{21} \equiv \hbar\gamma = 1.13~\mathrm{\mu eV}$ and $\hbar\gamma_{32} = 0.91~\mathrm{\mu eV}$  ($\mu = \sqrt{\gamma_{32}/\gamma_{21}} = 0.81$)~\cite{GerardotNJP2009}. As inferred from $\gamma_{21}$, $\mu_{21} = 0.6$~$e$ nm. 
The dielectric constant of the SQD is taken to be $\varepsilon_s = 13.9$. For the MNS, we chose the inner and outer radius to be  $r_1 = 9~\mathrm{nm}$ and $r_2 = 10~\mathrm{nm}$, respectively, which, according to Eq.~(\ref{alpha_1}), gives the energy of the LSP resonance $\hbar\omega_{\mathrm{LSP}} = 1.41~\mathrm{eV}$, 
which is around the energies of the ground--to--one-exciton and one-exciton--to--biexciton transitions as well as the peak position of the factor $|1 + \alpha_1(\omega_0)/(2\pi d^3)|$, Fig.~\ref{Enhancement}.
As a measure of the TPA efficiency, the population of the biexciton state $\rho_{33}$ is considered.

\subsection{Steady-state analysis}
\label{Steady-state analysis}
First, we examine the steady-state regime of the TPA setting the time derivatives in Eqs.~(\ref{rho11})--(\ref{rho31}) to zero. To solve the resulting system of nonlinear equations, we use the exact parametric method developed in Ref.~\onlinecite{RyzhovPRA2019} (see also Appendix A).
The stability of the steady-state solution is uncovered by making use of the standard Lyapunov exponent analysis~\cite{KatokBook1995}. To this end, we calculate the eigenvalues $\lambda_k$ ($k=1\ldots 8$) of the Jacobian matrix of the right hand side of Eqs.~(\ref{rho11})--(\ref{rho31}) as a function of $|\Omega|$. The exponent with the maximal real part, $\max_k{\mathrm{Re}\lambda_k}$, determines the stability of the steady-state solution: if $\max_k{\mathrm{Re}\lambda_k} \leq 0$ the solution is stable, while it is unstable otherwise.

\begin{figure*}
\includegraphics[width=0.8\linewidth]{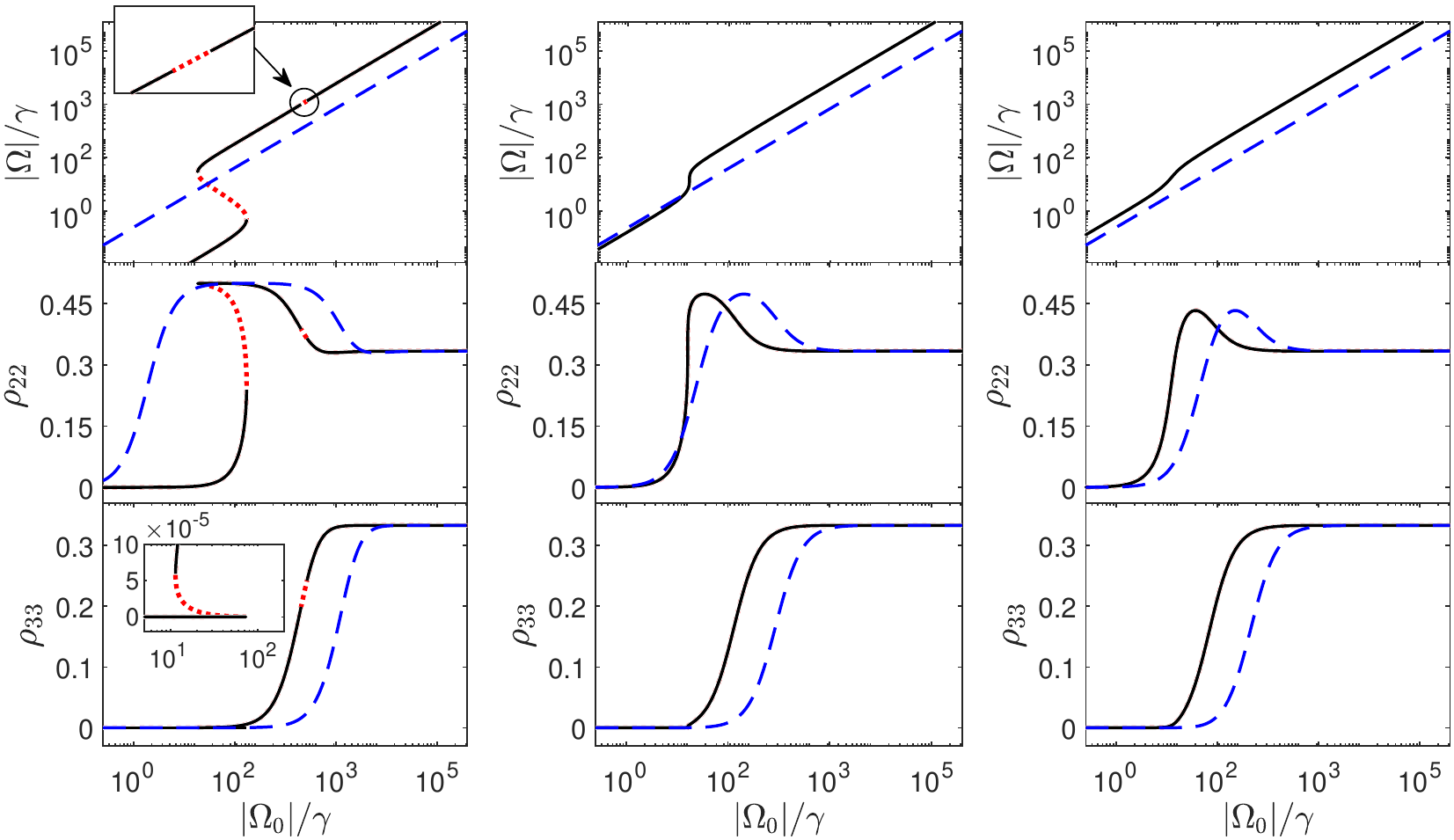}
\caption{\label{FieldDependentOneExciton} Steady-state solutions for the total field Rabi magnitude $|\Omega|$ and populations of the one-exciton and biexciton states, $\rho_{22}$ and $\rho_{33}$, respectively, as a function of the external field Rabi magnitude $|\Omega_0|$ calculated for the case of one-exciton resonance excitation ($\omega_0 = \omega_2$) for three values of the dephasing rates $\Gamma_2$ and $\Gamma_3$: left column -- $\Gamma_2 = \Gamma_3 =~\gamma$; middle column -- $\Gamma_2 = \Gamma_3 = 101.13\gamma$; right column -- $\Gamma_2 = \Gamma_3 = 300\gamma$. Solid (dashed) curves show the results obtained for the SQD-MNS hybrid setting the SQD-MNS center-to-center distance $d = 16~\mathrm{nm}$ (isolated SQD). The other parameters are described in the text. Dotted fragments of the curves indicate the unstable parts of the steady-state solutions.  The inserts blow up details of the curves.
}
\centering
\end{figure*}

Figure~\ref{FieldDependentOneExciton} illustrates the $|\Omega_0|$-dependence of the total field Rabi magnitude $|\Omega|$ and the populations of the one-exciton and biexciton states, $\rho_{22}$ and $\rho_{33}$, respectively, calculated for the case when the external field is in resonance with the one-exciton transition ($\omega_0 = \omega_2$). Three values of the dephasing rates $\Gamma_2$ and $\Gamma_3$ where considered : $\Gamma_2 = \Gamma_3 =~\gamma$ (left column), $\Gamma_2 = \Gamma_3 = 101\gamma$ (middle column), and $\Gamma_2 = \Gamma_3 = 300\gamma$ (right column). In the calculations, the SQD-MNS center-to-center distance was chosen to be $d = 16$~nm. For this value, the feedback parameter is found to be $\hbar G = (1.23 + 0.28~i)~\mathrm{meV}=(1.09 + 0.25~i)\times 10^3\hbar\gamma$, i.e. of the same order as $\Delta_B$. The results are presented by solid curves. For comparison, shown by the dashed curves are the results of similar calculations for an isolated SQD.

From Fig.~\ref{FieldDependentOneExciton}, we observe that the system's response, first, exhibits bistability which disappears upon increasing the dephasing rates, with $\Gamma_2 = \Gamma_3 = 101.13\gamma$ being the threshold for bistability to break down (middle column). The dot-marked  branch with negative slop is unstable.
Second, within the range of existence of bistability, the biexciton state is almost unpopulated. This is because,
due to the destructive interference of the external and secondary fields, 
the Rabi magnitude $|\Omega|$ is small, namely $|\Omega| \ll \Delta_B$. The biexciton population becomes notable and large compared to that of an isolated SQD (enhancement effect) in the pre-saturation regime, $|\Omega| \sim \Delta_B$, occurring around $|\Omega_0| \sim 100\gamma$. In the deep saturation regime, $|\Omega| \gg \Delta_B$, no enhancement of the TPA is observed.

Finally, in a narrow interval of changing $|\Omega_0|$ (shown in the insert), the steady-state regime is again unstable (left column). The character of this instability will be discussed in Sec.~\ref{Dynamics}.

\begin{figure*}
\includegraphics[width=0.8\linewidth]{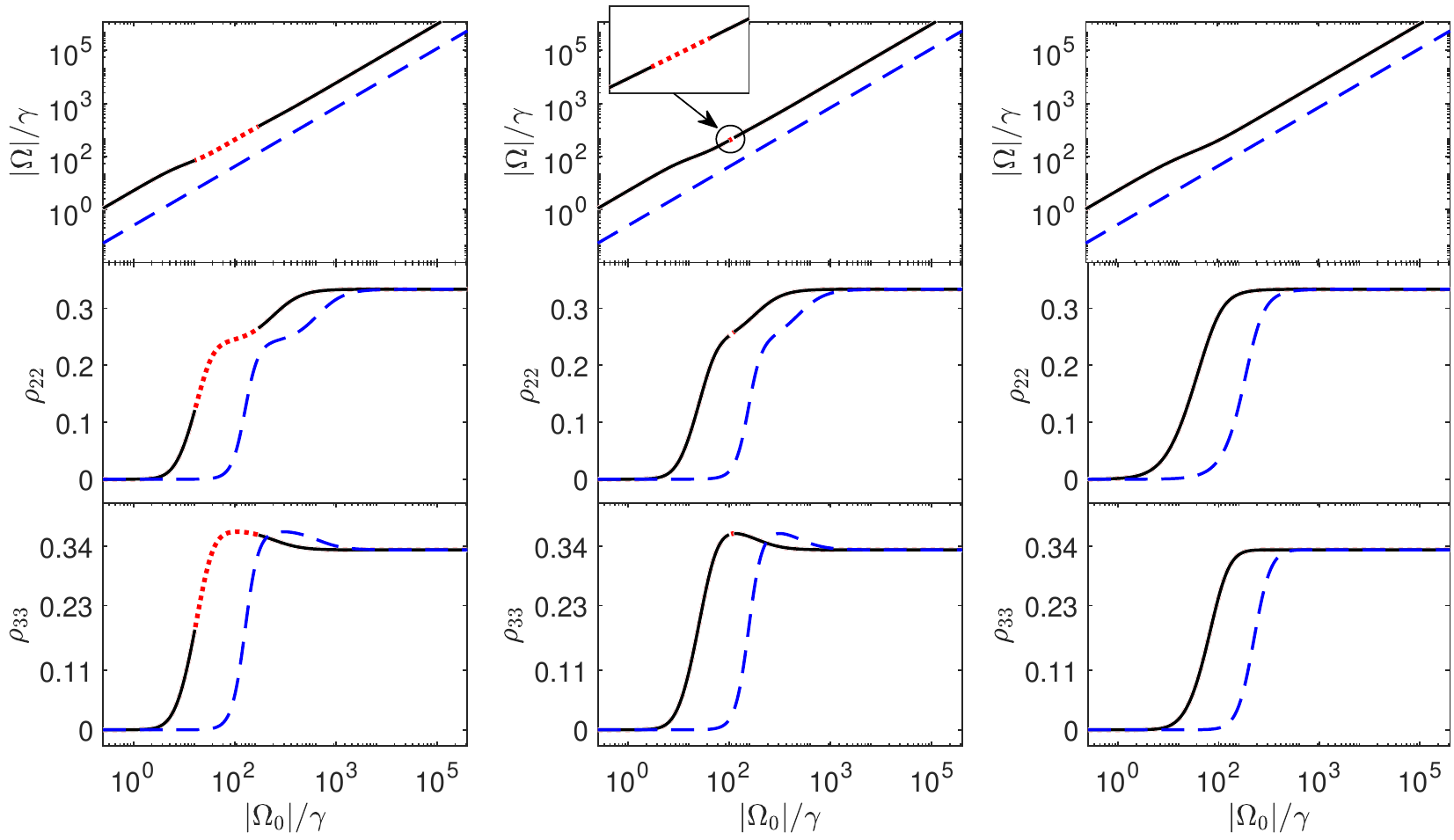}
\caption{\label{FieldDependentTPAresonance} Same as in Fig.~\ref{FieldDependentOneExciton}, but for the case of two-photon resonance excitation ($\omega_0 = \omega_3/2$) and three other values of the dephasing rates $\Gamma_2$ and $\Gamma_3$: left column -- $\Gamma_2 = \Gamma_3 = \gamma$; middle column -- $\Gamma_2 = \Gamma_3 = 4.6\gamma$; right column -- $\Gamma_2 = \Gamma_3 = 100\gamma$.
}
\centering
\end{figure*}

In Fig.~\ref{FieldDependentTPAresonance}, we present the results for the same quantities, but now calculated assuming that the external field is tuned to the two-photon resonance, $\omega_0 = \Omega_3/2$. In contrast with the previous type of excitation ($\omega_0 = \omega_2$), the response is single-valued within the whole range of the external field Rabi magnitude $|\Omega_0|$ and dephasing rates $\Gamma_2$ and $\Gamma_3$ considered. However, for $\Gamma_2 = \Gamma_3 = \gamma$ (left column), there exists a wide range of $|\Omega_0|$, where the system is unstable. This region shrinks upon increasing $\Gamma_2$ and $\Gamma_3$ and for $\Gamma_2 = \Gamma_3 > 4.7\gamma$ it disappears (see the middle column). Also we observe a peak value in the overall drastic enhancement of the TPA
within approximately a range of $|\Omega_0|\in [10,100]\gamma$, before the the transitions become saturated.

\subsection{Dynamics}
\label{Dynamics}
\begin{figure*}[ht]
\includegraphics[width=0.8\linewidth]{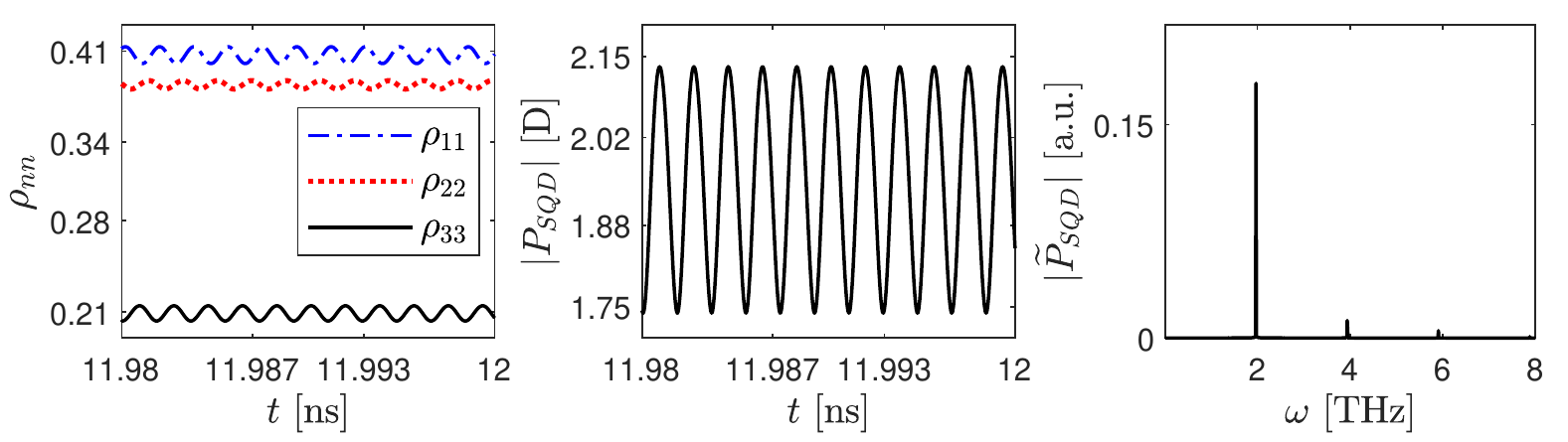}
\caption{\label{RhoVsT-1} Sustained dynamics (after transients are gone) of the SQD populations $\rho_{nn}(t)$ (n = 1,2,3) (left panel) and the SQD dipole moment magnitude $|P_{SQD}(t)| = \mu_{21}|\rho_{21}(t) + \mu\rho_{32}(t)|$ (middle panel) of the SQD-MNS composite calculated for the case when the external field is in resonance with the one-exciton transition ($\omega_0=\omega_2$). Right panel -- the Fourier spectrum of $P_{SQD}(t)$. Calculations where performed for the set of parameters of Fig.~\ref{FieldDependentOneExciton} (left column) at $|\Omega_0| = 615\gamma$ residing within the instability region shown in the insert in Fig.~\ref{FieldDependentOneExciton} (left column, upper plot).} 
\centering
\end{figure*}

\begin{figure*}[ht]
\includegraphics[width=0.8\linewidth]{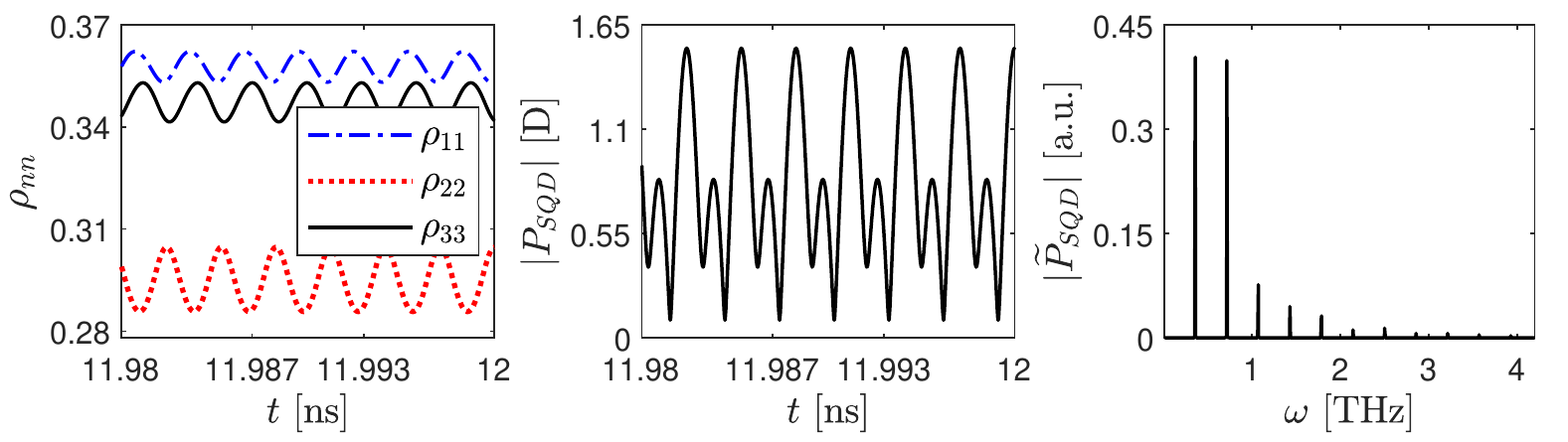}
\caption{\label{RhoVsT-2} Same as in Fig.~\ref{RhoVsT-1}, but but for the case of two-photon resonance excitation ($\omega_0 = \omega_3/2$) and the set of parameters of Fig.~\ref{FieldDependentTPAresonance} (left column) at $|\Omega_0|=85~\gamma$ residing within the instability region.
}
\centering
\end{figure*}

As is deduced from the steady-state analysis, there are windows of instability in the TPA of the SQD-MNS composite. In this section, we explore the nature of the TPA instabilities. To this end, we solve the dynamic equations~(\ref{rho11})--(\ref{rho31}) and~(\ref{Omega}), considering the SQD initially in the ground state [$\rho_{11}(0) = 1$] for a given external field Rabi magnitude $|\Omega_0|$ within the instability window (specified in the figure captions). The results of calculations are shown in Figs.~\ref{RhoVsT-1} and~\ref{RhoVsT-2}, which were obtained for two conditions of excitation: Fig.~\ref{RhoVsT-1} -- the external field is in resonance with the one-exciton transition ($\omega_0 = \omega_2$) and Fig.~\ref{RhoVsT-2} -- the external field is in resonance with the two-photon transition ($\omega_0 = \omega_3/2$).

The left panel in each figure displays the population dynamics after the transient stage is gone, the middle panel - the dynamics of the SQD's mean dipole moment magnitude $|P_\mathrm{SQD}(t)|$, and the right panel -- the Fourier spectrum of the SQD's mean dipole moment, $|\widetilde{P}_\mathrm{SQD}(\omega)| = |\int dt \exp(i\omega t)P(t)|$ (only the positive-frequency part is shown). The dynamics in both cases looks like self-oscillations, which is confirmed by the signal's Fourier spectra, having a well defined discrete structure of equidistantly spaced harmonics. Thus, self-oscillations are the only type of instabilities exhibited by these InGaAs/GaAs SQD -- silica-Au MNS composite.

\section{Summary}
\label{Summary}
We conducted a theoretical study of the two-photon absorption of a composite comprising a semiconductor quantum dot and a metallic nanoshell, considering the SQD as a three-level ladder-like system with ground, one-exciton and biexciton states. The presence of a MNS nearby the SQD is found to have a large impact on the TPA of the composite due to two principal effects: (i) -- renormalization of the applied field amplitude and (ii) -- renormalization of the resonance frequencies and radiation relaxation rates of the quantum dot, both depending on the quantum dot level populations.
In the perturbation regime, when the the biexciton state is only slightly populated, each of these factors may give rise to both suppression and enhancement of the TPA as compared to the TPA of an isolated SQD. The resulting effect is determined by the complicated interplay of those factors.

The nonlinear regime of the TPA (where the biexciton state is significantly populated) was analyzed for a particular case of a resonantly tuned composite comprizing an InGaAs/GaAs SQD and a silica-Au MNS separated by a center-to-center distance $d = 16$~nm. We found that the TPA of this heterostructure experiences a drastic enhancement compared to the TPA of an isolated SQD prior the SQD transitions become saturated. This occurs independently of the type of excitation, either into the one-exciton resonance or into the two-photon resonance.

Two more effects were uncovered in our results for the TPA of the composite that no analog in the TPA of an isolated SQD: first -- bistability of the TPA under the excitation of the SQD into the one-exciton resonance and, second, -- the emergence of a self-oscilllation regime in the TPA, existing for both types of excitations, either into the one-exciton or two-photon resonance. Both effects were found to disappear upon increasing the dephasing rates of the excitonic transitions.

To conclude, we note that InGaAs/GaAs SQDs absorb light in the inftared. When conjugated with MNSs, which drastically enhance the SQD optical response, they might be considered as promising candidates for application in biosensing and optical imaging.

\acknowledgments
This work was supported by the Directorate General of Higher Education, Ministry of Research, Technology and Higher Education of Indonesia. B.S.N. acknowledges the University of Groningen for hospitality.

\begin{appendix}
\section{Solution of the steady-state problem}
\label{AppendixA}
The steady-state problem is governed by the following set of equations:
\begin{widetext}
\begin{subequations}
\label{all_steadyR}
\begin{equation}
    \gamma(Z_{21} - Z_{32}) + 3i \left(\Omega^* \rho_{21} - \Omega \rho^*_{21} \right) = -\gamma~,
\label{rho11_steady}
\end{equation}
\begin{equation}
    \mu\gamma Z_{32} - i \left(\mu \Omega^* \rho_{21} - \Omega^*\rho_{32} - \mu \Omega \rho^*_{21}
    +\Omega \rho_{32}^*\right) = 0 ~,
\label{rho33_steady}
\end{equation}
\begin{equation}
    i\Omega Z_{21} + \left(i\Delta_{21} + \frac{1}{2}\gamma + \Gamma_2 \right) \rho_{21} - i\mu \Omega^* \rho_{31} = 0~,
\label{rho21_steady}
\end{equation}
\begin{equation}
     \mu \Omega Z_{32} + \left[ i\Delta_{32} + \frac{1}{2} (\mu^2 + 1)\gamma + \Gamma_2 + \Gamma_3 \right] \rho_{32} + i\Omega^* \rho_{31} = 0~,
\label{rho32_steady}
\end{equation}
\begin{equation}
    i(\mu \Omega \rho_{21} - \Omega \rho_{32})- \left(i\Delta_{31} + \frac{1}{2}\mu^2\gamma + \Gamma_3\right) \rho_{31} = 0~,
\label{rho31_steady}
\end{equation}
\end{subequations}
\end{widetext}
where $\Omega$ is given by Eq.~(\ref{Omega}). The main steps towards solving exactly Eqs.~(\ref{rho11_steady})--(\ref{rho31_steady}) together with Eq.~(\ref{Omega}) are as folllows~\cite{RyzhovPRA2019}. Consider $\Omega$ in Eqs.~(\ref{rho11_steady})--(\ref{rho31_steady}) as a parameter. This system of {\it linear} equations can be solved analytically. Formally, let us write Eqs.~(\ref{rho11_steady})--(\ref{rho31_steady}) in a matrix form ${\bf M}(\Omega)R = R_0$, where the column vectors $R = (Z_{21},Z_{32},\rho_{21},\rho_{32},\rho_{31},\rho_{21}^*,\rho_{32}^*,\rho_{31}^*)^\mathrm{T}$ and $R_0 = (-\gamma,0,0,0,0,0,0,0)^\mathrm{T}$, while the matrix $\textbf{M}$ can be easily inferred from Eqs.~(\ref{rho11_steady})--(\ref{rho31_steady}) (we do not present its explicit form). The vector $R$ is found as $R = {\bf M}^{-1}(\Omega)R_0$, where the inverse matrix ${\bf M}^{-1}(\Omega)$ also is known explicitly. Afterwards, the solutions for $\rho_{21}(\Omega)$ and $\rho_{32}(\Omega)$ are used in Eq.~(\ref{Omega}) to find $\Omega$ and furthermore all the density matrix elements $\rho_{ij}$ (see Ref.~\onlinecite{RyzhovPRA2019} for detail).
\end{appendix}

%

\end{document}